\documentclass{pasa}%
\usepackage[authoryear]{natbib}
\bibpunct{(}{)}{;}{a}{}{,}
\usepackage{amssymb,amsmath}
\usepackage{bmpsize}
\usepackage{float}
\usepackage{graphicx}
\usepackage{color}
\usepackage{amsmath}
\usepackage[toc,page]{appendix}
\usepackage{caption}
\usepackage{subcaption}
\usepackage{enumitem, color, amssymb}
\usepackage{upquote}
\usepackage{color}

\newcommand{\msun}{{\rm M}_{\odot}}
\newcommand*{\rom}[1]{\expandafter\@slowromancap\romannumeral #1@}

\begin{document}%

\title[On the primordial specific frequency of globular clusters]{On the primordial specific frequency of globular clusters in dwarf and giant elliptical galaxies}

\author[Ahmed H. Abdullah et al.]{Ahmed H.~Abdullah$^{1,2}$\thanks{Email:vipahmed.hasan@gmail.com}, Pavel Kroupa$^{3,5}$\thanks{Email:pkroupa@uni-bonn.de}, Patrick Lieberz$^3$ and Rosa Amelia Gonz\'{a}lez-L\'{o}pezlira$^{4}$\\
\affil{$^1$Department of Astronomy, College of Science, University of Baghdad, Baghdad,10071, Iraq}
\affil{ $^2$Argelander Institut f\"{u}r Astronomie der Universit\"{a}t Bonn, Auf dem H\"{u}gel 71, D-53121 Bonn, Germany}
 \affil{$^3$Helmholtz-Institut f\"{u}r Strahlen- und Kernphysik (HISKP), 
Universit\"{a}t Bonn, Nussallee 14-16, D-53115 Bonn, Germany}
\affil{$^4$Instituto de Radioastronomia y Astrofisica, UNAM, Campus Morelia, Michoac\'{a}n, Mexico, C.P.58089}
\affil{$^5$Charles University in Prague, Faculty of Mathematics and Physics, Astronomical Institute, V  Hole\v{s}ovi\v{c}k\'ach 2 CZ-18000\\ Praha Czech Republic}
}%

\label{firstpage}


\begin{abstract}

  Globular clusters (GC) are important objects for tracing the early evolution of a galaxy. We study the relation between the properties of globular cluster systems - as quantified by the GC specific frequency ($S_{N}$) - and the properties of their host galaxies. In order to understand the origin of the relation between the GC specific frequency ($S_{N}$) and galaxy mass, we devise a theoretical model for the specific frequency ($S_{N,th}$). GC erosion is considered to be an important aspect for shaping this relation, since observations show that galaxies with low densities have a higher $S_{N}$, while high density galaxies have a small $S_{N}$. We construct a model based on the hypothesis that star-formation is clustered and depends on the minimum embedded star cluster mass ($\rm {M}_{ecl,min}$), the slope of the power-law embedded cluster mass function ($\beta$) and the relation between the star formation rate (SFR) and the maximum star cluster mass ($\rm {M}_{ecl,max}$). We find an agreement between the primordial value of the specific frequency ($S_{Ni}$) and our model for $\beta$ between 1.5 and 2.5 with $\rm {M}_{ecl,min} \leqslant 10^{4}\msun$.

\end{abstract}

    \maketitle
%
%
\section{Introduction}
\label{sec:intro}
Globular clusters (GCs) are collisional stellar-dynamical near-spherical systems of $10^{4}-10^{7}$ stars and among the first stellar systems to form in the early Universe. GCs are found within  different morphological types of galaxies, from  irregular to  spiral and elliptical galaxies. Most of the GCs appear to have formed within a few Gyr after the Big Bang \citep{2003A&A...408..529G} and the properties of GC systems can be considered as important tracers for the formation and evolution of galaxies.

 One of the basic parameters to describe the globular cluster system of a galaxy is the specific frequency, $S_{N}$, which is the number of globular clusters, $N_{GC}$, divided by the $ V $-band luminosity of the galaxy, normalized at an   absolute magnitude of the galaxy in the $ V $-band ($M_{V}$) of -15 mag \citep{1981AJ.....86.1627H}:

\begin{equation}\label{eq1}
 S_{N}\equiv N_{GC}\times 10^{0.4(M_{V}+15)}. \end{equation}
The specific frequency measures the richness of a GC system and varies between galaxies of different morphological types: $S_{N}$ is smaller in late-type spiral galaxies than in early-type elliptical (E) galaxies \citep[e.g][]{1998ApJ...508L.133M}. Spiral galaxies have a $S_{N}$ between 0.5 and 2 \citep{2003MNRAS.343..665G,2004ApJ...611..220C,2007AJ....134.1403R}. For more luminous elliptical galaxies, $ S_N$ ranges from about  2 to 10  and tends to increase with luminosity, while $ S_{N}$ increases from a few to several dozen with decreasing galaxy luminosity for dE galaxies that posses GCs \citep{2007ApJ...670.1074M,2008ApJ...681..197P,2010MNRAS.406.1967G}. This difference in $ S_N$ between types of galaxies needs to be understood in terms of formation models of galaxies \citep{2002MNRAS.333..383B}.

The relation between $ S_{N}$ and total mass of a galaxy ($\rm {M}_{b}$) reveals a `U' -shape,  i.e.,  higher $ S_{N}$ for dwarfs galaxies and giant ellipticals (low and high- mass end of the scale, respectively), with a minimum $ S_{N}$ for galaxies at an intermediate mass, as shown in Figure (\ref{Fig1}) \citep{2013ApJ...772...82H}. 

There are many suggestions to explain  the  observed `U' -shaped relation between  $  S_{N}$ and $ M_{V}$ of the host galaxy. \cite{1982AJ.....87.1465F} proposed a tidal stripping model of GCs from smaller galaxies to explain the increasing value of $ S_{N}$ in cD galaxies (dominant ellipticals in the centers of clusters) and studies of the GC system around cD galaxies also supported this scenario \citep{1997AJ....113.1652F,1997ApJ...483..745N}.   \cite{1987nngp.proc...18S} and \cite{1992ApJ...384...50A} suggested elliptical galaxies formed from the mergers of spiral galaxies and lead to GC formation with a high efficiency which reflects the low $S_{N}$ of giant spiral galaxies.

\cite{2010MNRAS.406.1967G} investigated the trend of increasing $ S_{N}$ above and below the absolute galaxy magnitude of $ M_{V} \approx - 20$ mag and explain this trend by a theoretical model of GC specific frequency as a function of host galaxy dark matter halo mass  with  a universal specific GC formation efficiency $ \eta $.  This is the total mass of GCs divided by  the mass of the host dark matter halo, irrespective  of galaxy morphology and which has a mean value of $\eta = 5.5\times 10^{-5}$. \cite{2013MNRAS.435.1536W} studied the apparent or phantom  virial mass ($\rm {M}_{vir}$) of dark matter halos in Milgromian dynamics. They found $ S_{N}$ and $\eta$ to be functions of $\rm {M}_{vir}$. The number of GCs and $\eta$ increase for $\rm {M}_{vir} >10^{12}\msun$ and decreases for $\rm {M}_{vir}\leq10^{12}\msun $. 

Another ansatz to explain this `U' -shape is for galaxies with a small and large mass to have been very inefficient at forming stars. \cite{2013ApJ...772...82H} suggested these galaxies formed their globular clusters before any other stars, then had a star formation shut off. Star formation is likely regulated by supernova feedback and virial shock-heating of the infalling  gas for low and massive galaxies  respectively, while intermediate mass galaxies have a maximum star formation efficiency \citep{2006MNRAS.368....2D}. 

GC destruction can be important for the relation between $ S_{N}$ and $ M_{V}$  \citep{2014A&A...565L...6M}. Tidal erosion together with dynamical friction on the stellar component in different galaxies could produce different GC survival fractions, which may explain the present-day dependence of $ S_{N}(M_{V})$ \citep[cf with][]{2014A&A...565L...6M}. 
\cite{2017A&A...606A..85L} argue that $ S_{N}$ is also consistently a strong function of metallicity.

In this paper, we present a model for the specific frequency of GCs. It is based on the notion that star formation occurs in correlated star formation events which arise in the density peaks in the molecular clouds that condense from the galaxy's interstellar medium (ISM). These are spatially and temporally correlated with scales $<$ 1 pc and formation durations $<$ 1 Myr and can also be referred to as being embedded clusters.
 
This paper is organized as follows: in Section 2 we review that $S _{N}$ is reduced through erosion processes suggesting $S_{N}(M_{V})$ to be nearly constant. In Section 3 we determine the star cluster mass function population time-scales and the most-massive-cluster -- SFR relation. The theoretical model of the specific frequency of globular clusters ($ S_{N,th}$) is then presented in Section 4, which is based on the notion that star clusters are the basic building blocks of a galaxy \citep{2005ESASP.576..629K}. Finally, Section 5 contains the  conclusion.

\section{ GC populations and tidal erosion }

The specific frequency of GCs ($S_{N}$) is an important tool to understand the evolution of galaxies \citep{1991ARA&A..29..543H,2006ARA&A..44..193B}.

In this work, the data is taken from the Harris catalogue \citep{2013ApJ...772...82H}. We selected elliptical galaxies with masses ranging between $ 10^{7} $ and $10^{13} \msun$ (with the exception of M32). These  masses are dynamical masses of the galaxy,  $\rm {M}_{b}= 4 \sigma_{e}^{2} R_{e} G^{-1}$,  
where $\sigma_{e}$ [pc Myr$^{-1}$] is the velocity dispersion, $R_{e}$ [pc] the effective half light radius, and $G$ is the gravitational constant [$ G\approx 4.43\times10^{-3}$ pc$^{3} \msun^{-1}$ Myr$^{-2}$]. We refer to these masses as total masses ($\rm {M}_{b}$), since  the putative dark-matter halo has a small  contribution to the mass within this  radius \citep[e.g.][]{2010ApJ...717..803G,2011MNRAS.411.1435T, 2013ApJ...772...82H,2013MNRAS.434.1964S}, and stellar remnants from a top-heavy integrated galactic stellar initial mass function (IGIMF) account for this contribution \citep{2013MNRAS.436.3309W}.

\begin{figure}
\begin{center}
\includegraphics[width=84mm]{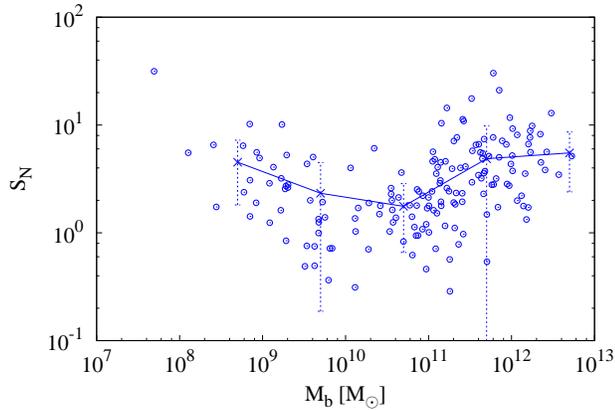}
\caption{The specific frequency of globular clusters versus total galaxy mass for a range of early-type galaxies using the data from \cite{2013ApJ...772...82H}. The crosses connected with a line are the average value of $ S_{N}$ per mass bin and the error bars are the standard deviation at the mass bin.} 
\label{Fig1}
\end{center}
\end{figure}

Figure (\ref{Fig1}) demonstrates a `U' -shape relation between $S_{N}$ (calculated using equation \ref{eq1}) and $\rm {M}_{b}$ \citep{2013ApJ...772...82H}. \cite{2014A&A...565L...6M} explained this relation as an effect of tidal erosion.
  For the purpose of understanding how tidal erosion contributes to this relation, one has to study the relation between the 3D mass density ($\rho_{3D}\equiv \rm {M}_{b}/{R_{H}}^3$) [$\msun/$pc$^{3} $] within the half-light radius (3D half mass radius $R_{H}$ $\approx$ 1.35 times the projected half light radius) and $\rm {M}_{b}$ [$\msun $].

The relation between the 3D mass density and the total mass takes the same trend as the $S_{N}$ vs. $\rm {M}_{b}$ relation as shown by \cite{2014A&A...565L...6M}. Near $\rm {M}_{b}$= $10^{10}\msun$ is the highest mean density, while the density is lower for less and more massive galaxies. Galaxies with different densities appear to  generate different GC survival times.
\begin{figure*}
\begin{center}
\includegraphics[width=150mm]{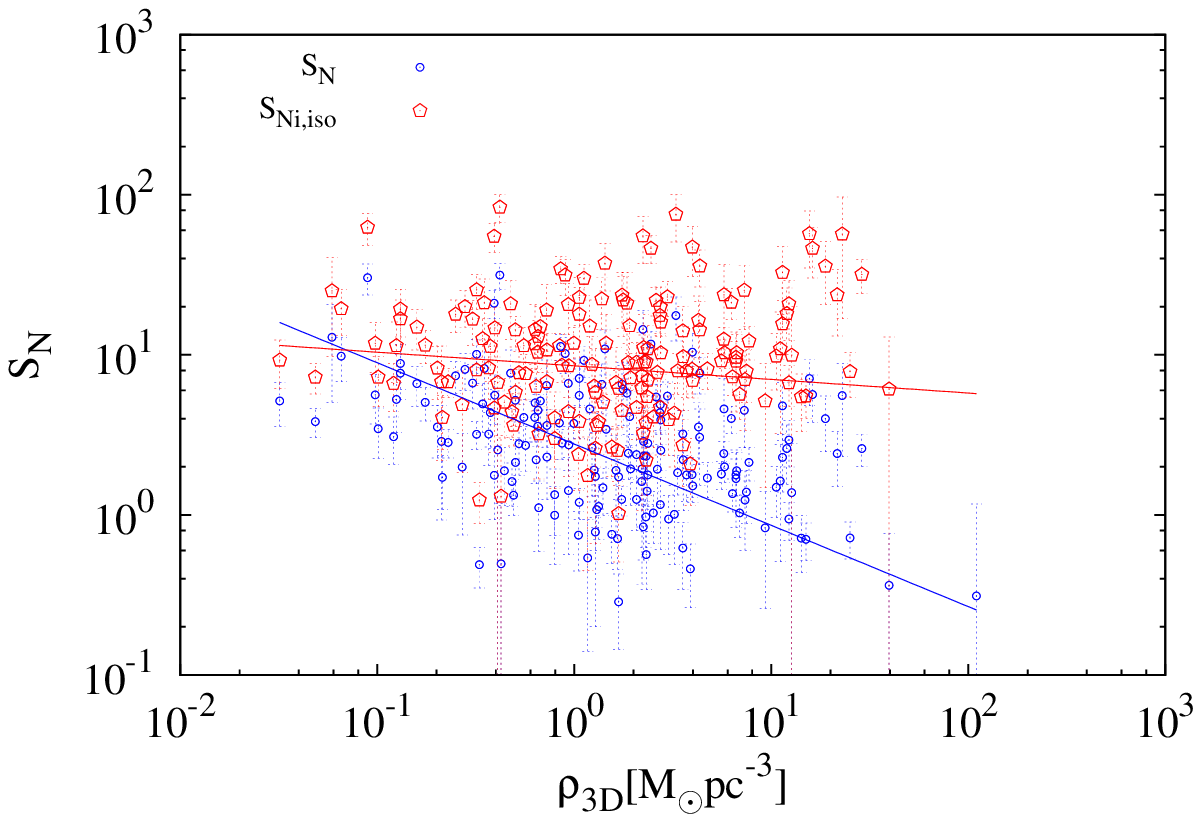}
\caption{The present and primordial values for specific frequencies of globular clusters versus the density ($\rho_{3D}$) of early-type galaxies. The red open pentagons ($S_{Ni,iso}$) are primordial values for $ S_{N}$ calculated for the isotropic case (equation \ref{eq5}). The blue circles are specific frequencies at the present epoch for the same sample as in Figure (\ref{Fig1}). The solid lines are the least square best-fits to the primordial and present cases by weighting with the error (dashed lines) in both directions.}
\label{Fig2}
\end{center}
\end{figure*} 
\cite{2014A&A...565L...6M} arrived at two equations to calculate the  GC survival fraction, $\textit{f}_{s}$, for initially isotropic ($\textit{f}_{s,iso}$) and radially anisotropic ($\textit{f}_{s,aniso}$) GC orbital velocity distributions respectively. After 10 Gyr of evolution, GCs more massive than $10^5 \msun$ will have:

\begin{equation}\label{eq2}
\textit{f}_{s,iso}= -0.160\times {\rm log_{10}}(\rho_{3D})+0.315, 
\end{equation}
\begin{equation}\label{eq3}
\textit{f}_{s,aniso}= -0.182\times {\rm log_{10}}(\rho_{3D})+0.216.
\end{equation}

According to equations (\ref{eq2}) and (\ref{eq3}), more GCs get destroyed at higher densities for an initially radially anisotropic velocity distribution function.

The correlation between the $ S_{N}$ and $ \rho_{3D}$ is determined observationally: Figure (\ref{Fig2}) shows the observed present-day GC specific frequency and $\rho_{3D}$ for the same sample as in Figure (\ref{Fig1}), which supports a high erosion of GCs at higher densities. The solid blue line is the bi-variate best fit to the observed data: 

\begin{equation}\label{eq4}
 S_{N}=(2.77 \pm 0.07)\times({\rho_{3D}})^{-0.51 \pm 0.02}.
\end{equation}

This supports the notion that the survival fractions $f_{s}$ of GCs may be an important aspect of the `U' -shaped relation between $ S_{N}$ and $\rm {M}_{b}$  as suggested by \cite{2014A&A...565L...6M}. 

In order to estimate the primordial value of the specific frequency, $ S_{Ni}$, for both cases, isotropic, $S_{Ni,iso}$, and radially anisotropic GC velocity distributions, 
$ S_{Ni,aniso}$, we divide $S_{N}$  by $\textit{f}_{s,iso}$ and by $\textit{f}_{s,aniso}$, respectively,

\begin{equation}\label{eq5}
 S_{Ni,iso}=\dfrac{S_{N}}{\textit{f}_{s,iso}},
\end{equation} 

\begin{equation}\label{eq6}
S_{Ni,aniso}=\dfrac{S_{N}}{\textit{f}_{s,aniso}}.
\end{equation} 
The primordial $S_{Ni,iso}$ and the observed present-day specific frequency at different densities are illustrated in Figure (\ref{Fig2}). As already concluded by \cite{2014A&A...565L...6M} it emerges that the initial specific frequency ($S_{Ni}$) is largely independent of $\rho_{3D}$. This result has potentially very important implication for our understanding of early galaxy assembly: \textit{ $S_{Ni}$ being nearly constant with density, the efficiency of forming young GCs (i.e. the number of young clusters per mass) is about the same for all present-day early- type galaxies from dEs to Es, suggesting that the same fundamental principle was active, independent of the mass of the galaxy.}

The primordial values of the number of globular clusters for the isotropic,  $N_{GCi,iso}$, and anisotropic, $N_{GCi,aniso}$, cases is calculated by dividing the observed number of globular clusters $N_{GC}$ by the GC survival fractions $\textit{f}_{s,iso}$ and $ \textit{f}_{s,aniso}$, respectively, 

\begin{equation}\label{eq7*}
N_{GCi,iso}=\dfrac{N_{GC}}{\textit{f}_{s,iso}},
\end{equation} 

\begin{equation}\label{eq8*}
N_{GCi,aniso}=\dfrac{N_{GC}}{\textit{f}_{s,aniso}}.
\end{equation} 

\begin{figure}
\begin{center}
\includegraphics[width=84mm]{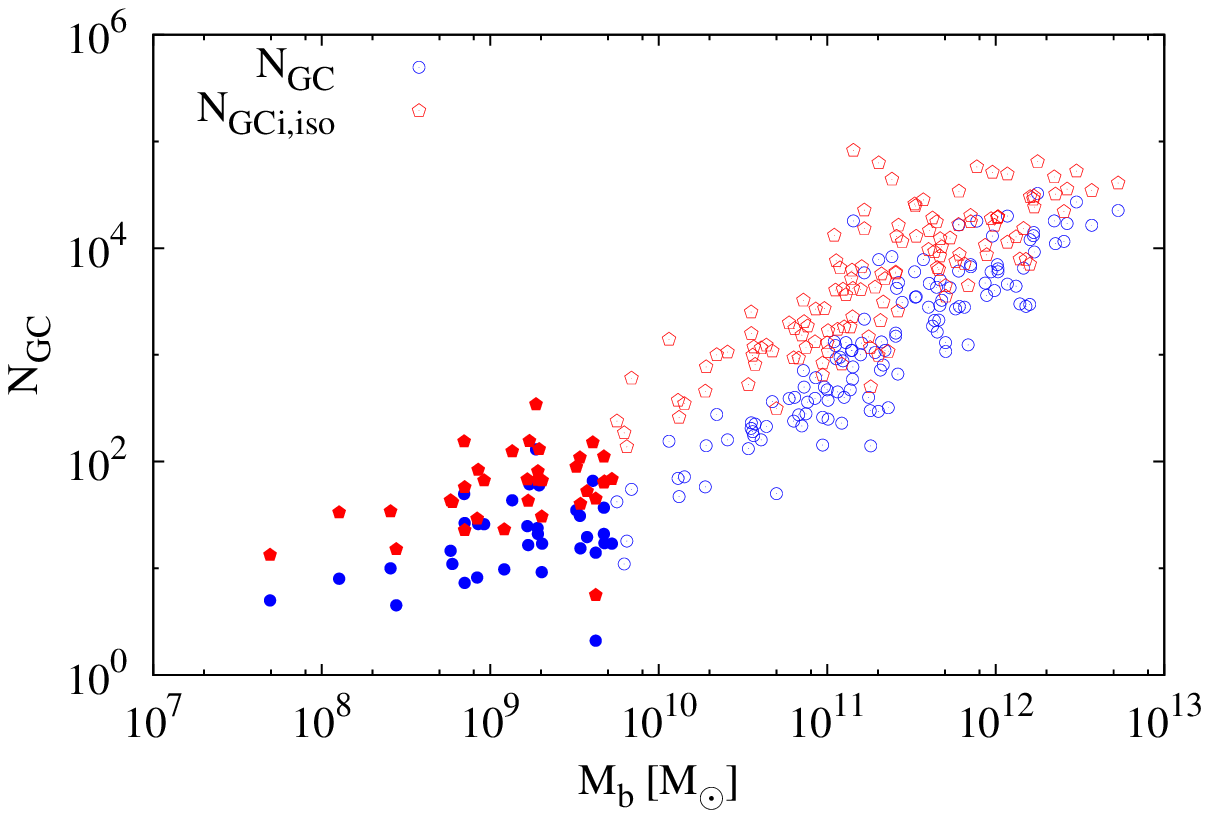}
\caption{The primordial number of GCs (red pentagons) for the isotropic case and the present-day  number of GCs (in blue points) versus the total mass of a galaxy. The solid symbols are  dE galaxies with stellar mass $< 5\times 10^{9} \msun$, while open symbols denote more-massive E-type galaxies.} 
\label{Fig3}
\end{center}
\end{figure}

The primordial number of GCs increases monotonically with host galaxy mass. Figure (\ref{Fig3}) shows this relation for $N_{GCi,iso}$ (red pentagons) and $N_{GC}$ (blue circles) as a function of $\rm {M}_{b}$. Filled symbols are galaxies with a mass smaller than $5\times 10^{9}$ $\msun$  denoted by branch \rm{I} (B\rm{I}), while open symbols are galaxies with a mass larger than $5\times 10^{9}$ $\msun$ denoted by branch \rm{II} (B\rm{II}). Galaxies in branch \rm{I} are dEs, while E galaxies are in branch \rm{II} \citep{2008MNRAS.386..864D}. The present-day number of GCs, $N_{GC}$, is lower than $N_{GCi,iso}$, especially for galaxies at intermediate-mass, because of the high destruction rates of GCs. 

We now make the following ansatz: if the fundamental physical processes acting during the assembly of dE and E galaxies were the same, the former formed fewer GCs because their SFRs were much smaller than during the formation of E galaxies \citep{2004MNRAS.350.1503W,2013ApJ...775L..38R,2013MNRAS.436.3309W}. This ansatz is followed through  a model in the next section.

\section {The {$\rm {M}_\MakeLowercase{ecl,max}$} - $SFR$ correlation and the star cluster mass function population time-scale (\MakeLowercase{$ \delta t$)}}

An empirical relation has been derived by \cite{2009A&A...499..711R} for dE and E galaxies between the central velocity dispersion $\sigma$ [km/s], which reflects the total stellar mass, and the stellar alpha-element abundance [$\alpha$/Fe]. The implied star formation duration, $ \Delta T$ [yr], over which the galaxy assembled, inversely correlates with the mass $\rm {M}_{b}$ [$\msun$] of the galaxy \citep[this is referred to as downsizing, see also][]{1999MNRAS.302..537T},

\begin{equation}\label{eq7}
\Delta T=10^{(11.38-0.24 \times {\rm log_{10}} (\rm {M}_{b}) )}.
\end{equation}

\begin{figure}
\begin{center}
\includegraphics[scale=0.65, angle=0]{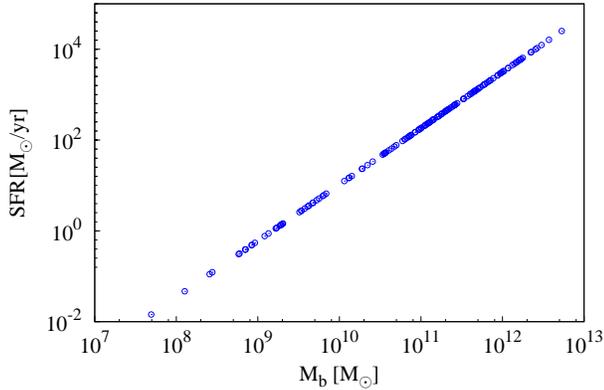}
\caption{The star formation rates (SFR) as a function of $\rm {M}_{b}$, with the  SFR from the $ \Delta T$-mass relation of \cite{2009A&A...499..711R} (their equation (19)). The  $\rm {M}_{b}$ values are taken from the Harris catalogue  \citep{2013ApJ...772...82H}.}
\label{Fig4}
\end{center}
\end{figure}

Knowing the total mass of a galaxy, $\rm {M}_{b}$, and $ \Delta T$,  the star formation rate (SFR) follows as illustrated in Figure (\ref{Fig4}),

\begin{equation}\label{eq8}
SFR =  \dfrac{\rm {M}_{b}}{\Delta T}.
\end{equation}

The time during which GCs formed ($\Delta t_{1} $) is part of the time scale of star formation in galaxies ($\Delta T $), $\Delta t_{1} \leq \Delta T$. Each part is divided into star cluster-population formation epochs of equal length $\delta t$, which we will calculate later (Figure \ref{Fig6}).
Assuming that all the stars form in embedded
 star clusters \citep{2003ARA&A..41...57L,2005ESASP.576..629K,2016AJ....151....5M}, the total mass of the star cluster system formed during $\delta t$ ($\rm {M}_{tot,\delta t}$), can be calculated using the SFR and $\delta t$, 

\begin{equation}\label{eq9}
\rm {M}_{tot,\delta t}=SFR \times\delta t.
\end{equation} Observational studies suggest that the masses of young and embedded star-clusters are distributed as a power law:

\begin{equation}\label{eq10}
\xi_{ecl}(\rm {M}_{ecl})= \rm {K}_{ecl} \left({\dfrac{{\rm {M}_{ecl}}}{{\rm {M}_{ecl,max}}}}\right)^{-\beta},
\end{equation} where $\xi_{ecl}$ is the mass distribution function of the embedded clusters, $\rm {K}_{ecl}$ is a normalization constant and $\rm {M}_{ecl}$ is the stellar mass of the embedded cluster. The power law slope $\beta$ is found to be between 1.2 and 2.5 ~\citep{1997ApJ...480..235E,2003ARA&A..41...57L,2003ApJ...598.1076K,2004MNRAS.350.1503W,2010AJ....140...75W,0004-637X-727-2-88}.
The total mass of a population of star clusters, $\rm {M}_{tot,\delta t}$, assembled within the time span $\delta t$ can also be expressed as follows
\begin{equation}\label{eq11}
\rm {M}_{tot,\delta t}=\int_{\rm {M}_{min}}^{\rm {M}_{ecl,max}(SFR)} \xi_{ecl}(\rm {M}_{ecl}) \rm {M}_{ecl} \rm {dM}_{ecl}, 
\end{equation}
where $\rm {M}_{min}$ is the minimum mass of a star cluster and $\rm {M}_{ecl,max}$ is the maximum star cluster mass depending on the SFR \citep{2004MNRAS.350.1503W}. $\rm {M}_{min}$ can be assumed to be 5 $\msun$, which is about the lowest  mass cluster observed to form in the nearby Taurus-Auriga  aggregate ~\citep{2002ApJ...580..317B,2003MNRAS.346..369K,2004MNRAS.350.1503W}. 

\begin{figure}
\begin{center}
\includegraphics[scale=0.40, angle=0]{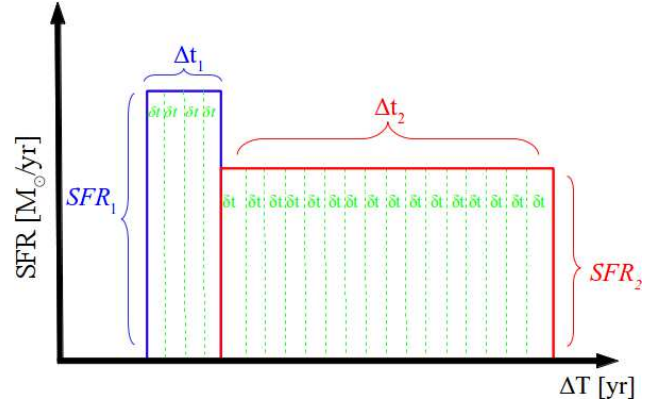}
\caption{Schematic drawing of the duration $\Delta T$ from \cite{2009A&A...499..711R}(divided into formation epochs of length $\delta t$) and the star formation rates (SFR) for the whole galaxy. SFR$_{1}$ is the SFR of forming the GC system over time $\Delta t_{1}$  and the rest of galaxy forms with SFR$_{2}$ over time $\Delta t_{2}$. $\Delta t_{1}$ is plotted here as preceding $\Delta t_{2}$ for illustrative purpose only.}
\label{Fig5}
\end{center}
\end{figure}

In order to determine the normalization constant $\rm {K}_{ecl}$ in equation (\ref{eq10}) we use the same assumption as in \cite{2004MNRAS.350.1503W}, that $\rm {M}_{ecl,max}$ is the single most massive cluster formed in time $\delta t$. For $\beta > 1$ and $\beta\neq 2$ (equation \ref{eq13} undefined) we get

\begin{figure*}
\begin{center}
\includegraphics[width=150mm]{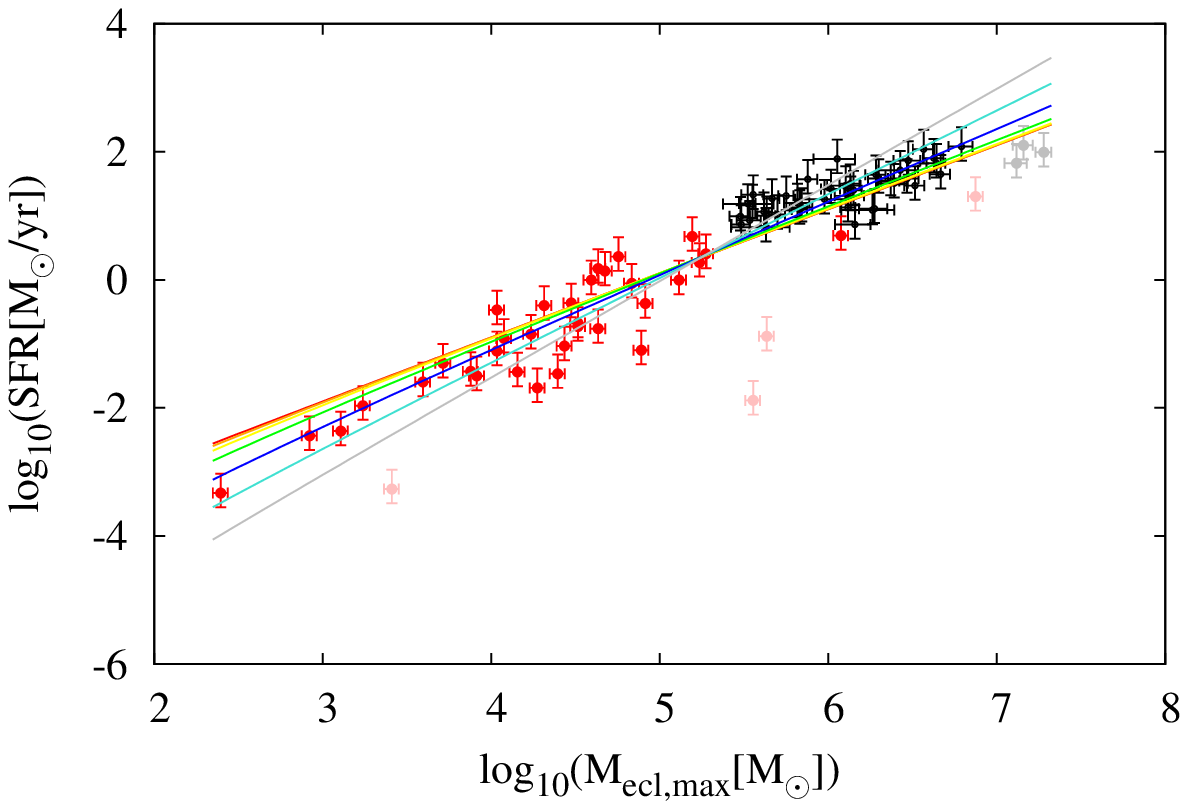}
\input{file-key.tex}
\caption{The cluster-system mass function population time-scale, $\delta t$, is determined by fitting the $SFR$ - $\rm {M}_{ecl,max}$ relation from equation (\ref{eq14}) to all data points using a weighted least-squares method. The best-fit $\delta t$  increases with increasing  $\beta$ (Table 1). The observational data (red circles) are taken from  \cite{2004MNRAS.350.1503W} with additional data points (black circles) from \cite{2013ApJ...775L..38R}. The faded color points are galaxies which were excluded from the least-square fits (see Section 3 for details).}
\label{Fig6}
\end{center}
\end{figure*}
\begin{table}
\caption{Time scale on which the embedded star cluster mass function population builds up, $\delta t$, for different embedded cluster mass function slopes ($\beta$). $\chi^{2}_{red}$ is extracted from the fits in Figure (\ref{Fig6}). The number of data points used is 73 (Figure \ref{Fig6}) with two free parameters ($\beta$ and $\delta t$) }
\label{my-label}
\begin{center}
\begin{tabular}{|c||c||c|}

\multicolumn{1}{|l|}{$\beta$} & \begin{tabular}[c|]{@{}c@{}}$\delta t$ \\   {[}$10^{6}$yr{]}\end{tabular} & \multicolumn{1}{l|}{$\chi^{2}_{red}$} \\ \hline
  1.2 &  0.08 & 1.78 \\ 
 
  1.5 &  0.16 & 1.76 \\ 
 
  1.7 & 0.29 & 1.69 \\ 
  
  1.9 & 0.69 & 1.53 \\ 
  
  2.1 &  2.30 & 1.47 \\ 
  
  2.3 & 11.00 & 1.33\\ 
  
  2.5 & 64.20 & 2.33 \\ 
  \hline
\end{tabular}
\end{center}
\end{table}
\begin{equation}\label{eq12}
\rm {K}_{ecl}=\dfrac{\beta-1 }{\rm {M}_{ecl,max}} \end{equation} 
and equation (\ref{eq11}) becomes, 

\begin{equation}\label{eq13}
 \rm {M}_{tot,\delta t}=\rm {M}_{ecl,max}^{\beta-1} \left(  \rm {M}_{ecl,max}^{2-\beta}-\rm {M}_{min}^{2-\beta} \right)\frac{\beta-1}{2-\beta}.
\end{equation} In order to determine $\rm {M}_{ecl,max}$ we correlate the theoretical upper mass limit of the star clusters and the most massive star cluster, using the same criteria as \cite{2015A&A}, which requires only one most massive cluster to exist ($1 =\int_{\rm {M}_{ecl,max}}^{\rm {M}_{ecl,max*}} \xi_{ecl}(\rm {M}_{ecl}) \rm {dM}_{ecl}$, where the theoretical upper
mass limit, ($\rm {M}_{ecl,max*}$) $ \geq \rm {M}_{ecl,max}$). 

According to the conditions above and by combining equation (\ref{eq9}) and (\ref{eq13}) we obtain a relation between the SFR and $\rm {M}_{ecl,max}$ for $\beta > 1$ and $\beta\neq 2$: 

\begin{equation}\label{eq14}
SFR =\frac{\rm {M}_{ecl,max}{S}^{-1}}{\delta t} \left(1- \left(\frac{\rm {M}_{min}S}{\rm {M}_{ecl,max} }\times\frac{\beta-1}{2-\beta}\right )^{2-\beta}\right),
\end{equation} with $S=(1-2^{\frac{2-\beta}{1-\beta}})$.

Observations indeed indicate that  young massive star clusters follow a relation between the visual absolute magnitude of the brightest young cluster and the global SFR of the host galaxy  \citep{2002AJ....124.1393L}. Based on this evidence \cite{2004MNRAS.350.1503W} found a relation between the galaxy-wide SFR and the maximum star-cluster mass. As indicated in equation (\ref{eq9}) the total mass depends on the current SFR at a certain $\delta t$, such that $\rm {M}_{ecl,max}$ depends on the SFR (equation \ref{eq14}), which has also been determined observationally \citep{2000A&A...354..836L,2004MNRAS.350.1503W}. It follows that galaxies with a high SFR are forming high-mass clusters.
\begin{figure}
\begin{center}
\includegraphics[scale=0.65, angle=0]{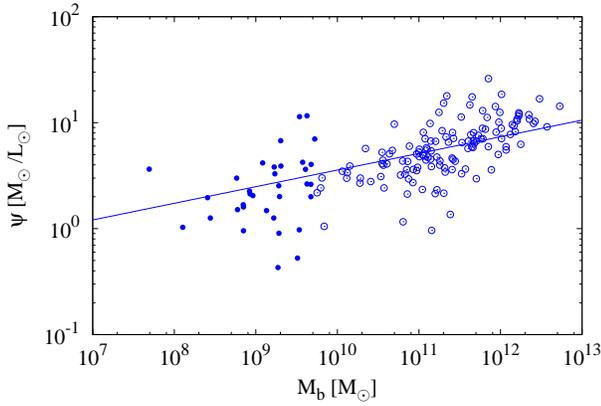}
\caption{The mass-to-light ratio, $\psi$, of the galaxies in the $V$- band as a function of $\rm {M}_{b}$, for the same sample as in Figure (1).}
\label{Fig7}
\end{center}
\end{figure}

The resulting relation between the SFR and the mass of the most massive cluster is illustrated in Figure (\ref{Fig6}). The data are from  \cite{2004MNRAS.350.1503W} and \cite{2013ApJ...775L..38R}. \cite{2013ApJ...775L..38R} provide $K$-band magnitudes which we converted to $V$-band magnitudes using the colour-magnitude relation $V-K = 2$ \footnote{\cite{2013ApJ...775L..38R} use this relation to convert the data from \cite{2004MNRAS.350.1503W} to the $K$-band. It is also roughly the colour of a $10^{7}$ yr old population with solar metallicity \citep{2003MNRAS.344.1000B}.}. We converted the luminosity of the brightest star cluster in the $V$-band to the most massive star-cluster mass using equation (5) from \cite{2004MNRAS.350.1503W}. We used these data to determine the length of the formation epoch $\delta t$. The uncertainties in SFR were obtained from the uncertainties in conversion of the IR luminosity to a SFR. On the other hand the uncertainties in $\rm {M}_{ecl,max}$ come from uncertainties in the conversion of luminosities to masses. 

We exclude seven galaxies (faded colors) in Figure (\ref{Fig6}) from this population: The first four galaxies (in increasing $\rm {M}_{ecl,max}$ values) are excluded since the SFRs of these dwarf galaxies do not represent the birth of these clusters \cite[further details can be found in][]{2004MNRAS.350.1503W,2015A&A}. The last three galaxies (gray) have a luminosity distance from the NED database larger than 150 Mpc. \cite{2013ApJ...775L..38R} suggested that these brightest super star clusters might be a blend of many clusters.  

The cluster-system mass function population time-scale, or the duration of the star formation ‘epoch’, $\delta t$, is determined by fitting equation (\ref{eq14}) for $\beta$ = 1.2 - 2.5 to the data (Figure \ref{Fig6}). The best value of $\delta t$ as a function of  ${\beta}$  is determined  by the reduced chi-squared statistic $\chi^{2}_{red}$. 
As can be seen in Table 1 and Figure (\ref{Fig6}), $\delta t$ increases with ${\beta}$. This result agrees with  \cite{2015A&A}, who found by comparison with the literature that ${\beta}$ lies between 1.8 and 2.4. Also it is consistent with the analysis by \cite{2004MNRAS.350.1503W}. $\chi^{2}_{red}$ is minimized for ${\beta}=2.3$, $\delta t=10^{7}$yr as already noted by \cite{2004MNRAS.350.1503W}. 
The typical time-scale of about $10^{7}$yr has also been deduced from calculations of the Jeans time in molecular clouds \citep[e.g.][]{2004PASJ...56L..45E}. The star formation time-scale can also be determined from examining offsets between $H\alpha$ and $CO$ arms of a spiral galaxy as proposed by \cite{2009ApJ...697.1870E}, who found the  star formation time to be between 4 and 13 Myr. Independently of these arguments, \cite{2010ARA&A..48..547F} review molecular cloud formation and find that on a time scale of 20-30 Myr the interstellar medium completes a cycle through the molecular phase with embedded star formation. This time scale is verified by \cite{2015ApJ...806...72M}. These time scale constraints are well consistent with $\delta t\approx 10$ Myr required to best match the data in Figure (\ref{Fig6}). We thus assume that every $\delta t\approx 10$ Myr a new population of star clusters hatches from the ISM of a star forming galaxy, follow the embedded cluster mass function (ECMF).

\section {Theoretical specific frequency {$(S_{N,\MakeLowercase{th}})$}}
With equation (\ref{eq1}), we can now derive an analytical model for $S_{N,th}$, which is the theoretical number of globular clusters, $N_{GC,th}$, per unit galaxy luminosity in the V-band. The galaxy luminosity can be converted into a mass such that:

\begin{equation}\label{eq15}
S_{N,th}=\frac{N_{GC,th}}{\rm {M}_{b}} \times \psi 10^{7.9} L_{V{\odot}},
\end{equation} where $\psi$ is the stellar mass-to-light ratio of the galaxy in the appropriate photometric band. Figure (\ref{Fig7}) shows  $\psi$ for the photometric band by using the same sample as in Figure (\ref{Fig1}). The  best least square fit suggests, 
\begin{equation}\label{eq16}
\psi= a\left(\frac{\rm {M}_{b}}{10^{6}\msun}\right)^{b} \frac{\msun}{L_{V{\odot}}} ,
\end{equation}
with a = 0.80 $\pm$ 0.13 and b = 0.15  $\pm$ 0.01.
\begin{figure*}
    \centering
    \begin{subfigure}[]{0.1\textwidth}
        \centering
        \includegraphics[scale=0.65, angle=0]{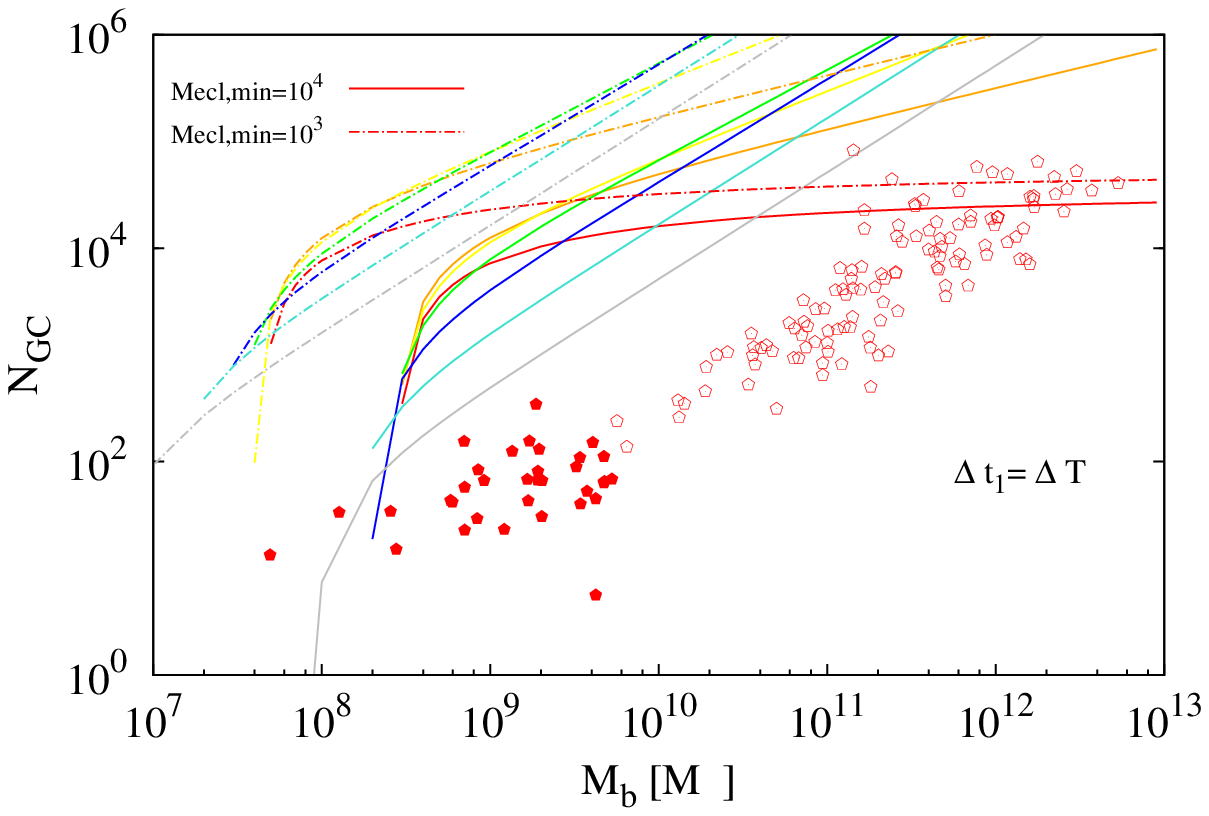}
        \caption*{}
        \label{fig:1*}
    \end{subfigure}
    \hfill
    \begin{subfigure}[]{0.5\textwidth}
        \centering
        \includegraphics[scale=0.65, angle=0]{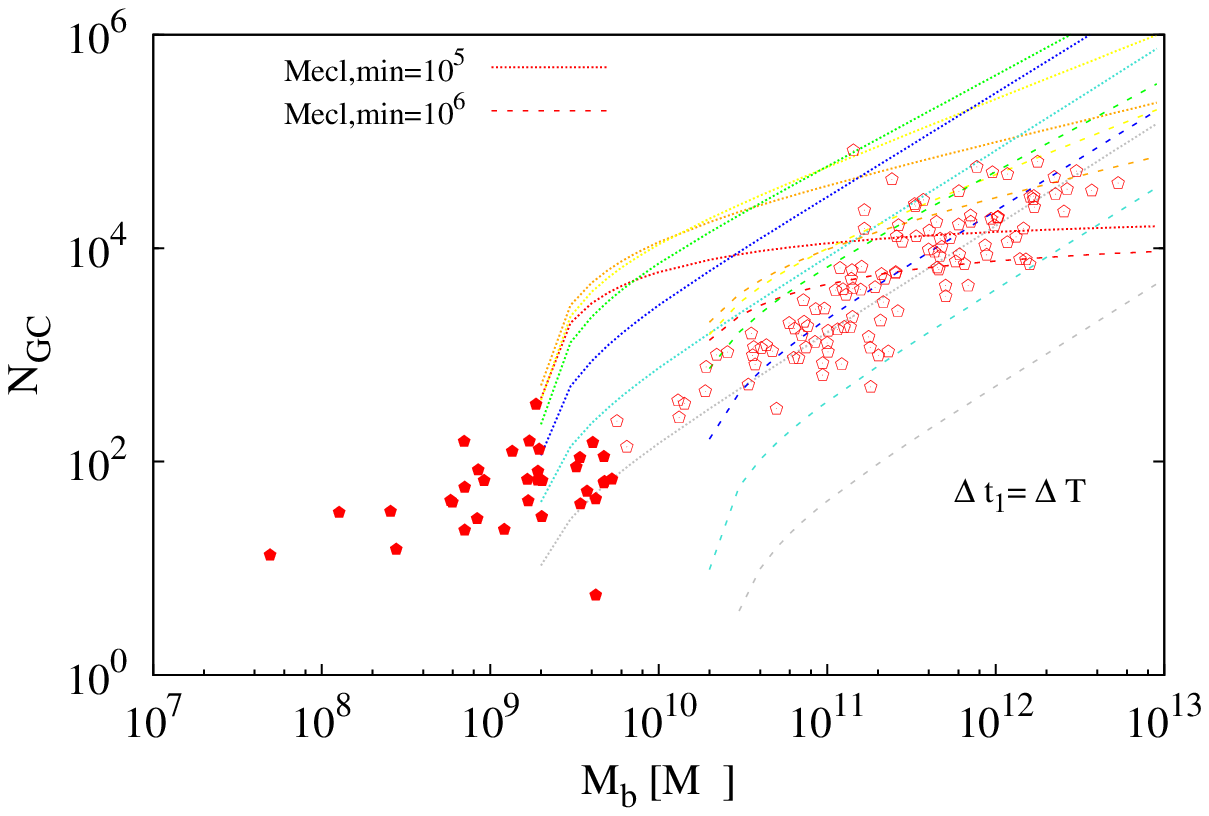}
        \caption*{}
        \label{fig:2*}
    \end{subfigure}
 \caption{Comparison between the primordial value (red pentagons) of the number of globular clusters, $N_{GCi,iso}$, and the theoretical number ($N_{GC,th}$) of globular clusters (coloured lines). Filled pentagons  are dE galaxies with masses $< 5 \times 10^{9}\msun$ (B\rm{I} galaxies) while open pentagons are E galaxies with masses $> 5 \times 10^{9}\msun$ (B\rm{II} galaxies). The coloured lines are our models for different $\beta$ of the ECMF ranging between 1.2 to 2.5 (red to gray as in Figure 6). In the left panel, we plot the primordial values for the number of globular clusters, $ N_{GCi,iso}$, and models for $\rm {M}_{ecl,min}=10^{3} \msun$ (dash-dotted lines) and $\rm {M}_{ecl,min}=10^{4}\msun$ (solid lines) for $\Delta t_{1}= \Delta T $. In the right panel we plot $N_{GCi,iso}$  and models with $\rm {M}_{ecl,min}=10^{5}\msun$ (dotted lines), $\rm {M}_{ecl,min}=10^{6}\msun$ (dashed lines) and for $\Delta t_{1}= \Delta T $.  Note that the SFRs of dE galaxies are too small to allow the formation of clusters with $\rm {M}_{ecl,min} \geqslant10^{5}\msun$ within the ansatz $\Delta t_{1}= \Delta T $ (Section 4.1).}    
    \label{Fig9}
\end{figure*}
In order to estimate $S_{N,th}$ from equation (\ref{eq15}), the number of globular clusters ($N_{GC,th}$) is required. The GCs which formed during  $\delta t$ can be calculated using the ECMF,

\begin{equation}\label{eq17} 
N_{GC}=\int_{\rm {M}_{ecl,min}}^{\rm {M}_{ecl,max}(SFR)} \xi_{ecl}(\rm {M}_{ecl}) \rm {dM}_{ecl}, 
\end{equation} to give

\begin{equation}\label{eq18}
\begin{aligned}
N_{GC,th} & =\frac{(\beta-1)\rm {M}_{ecl,max}^{\beta}}{(1-\beta) \rm {M}_{ecl,max}}\left[  \rm {M}_{ecl,max}^{1-\beta}-{\rm {M}_{ecl,min}}^{1-\beta}\right]\\
 & ~~ ~ \times \frac {\Delta t_{1}}{\delta t} .
\end{aligned}
\end{equation}
  Note the difference between $\rm {M}_{min}$ in equation (\ref{eq13}) and $\rm {M}_{ecl,min}$ in  equation (\ref{eq18}), since $\rm {M}_{min}$ is the physical lower limit for the embedded cluster mass \citep{2004MNRAS.348..187W}. From the SFR - $\rm {M}_{ecl,max}$ relation (equation \ref{eq14}) and to calculate  $\rm {M}_{ecl,max}$ we assume two cases of SFR: in case one the SFR is constant and equal over the time scales $\delta t$, $\Delta t_{1} $  and $\Delta t_{2} $; in case two the SFR is not equal in the time scale $\Delta t_{1} $ and $\Delta t_{2} $. While SFR$_{1}$ is constant for t=t$_{\circ}$ until t=t$_{\circ} + \Delta t_{1} $ and SFR$_{2}$ is constant for t=(t$_{\circ}+\Delta t_{1}$) until t=(t$_{\circ} + \Delta t_{1} $)$+\Delta t_{2}$. The minimum mass of clusters ($\rm {M}_{ecl,min}$) which become after a few Gyr globular cluster is assumed to be $10^{3}, 10^{4}, 10^{5}$ and $10^{6}\msun $. 
\subsection{Constant and equal SFR over $\delta t$, $\Delta t_{1} $ and $\Delta t_{2} $ }
In order to calculate the maximum mass of the old cluster systems in a galaxy , i.e., the clusters formed in the time span $\Delta t_{1}$, we assume that the young stars formed in embedded clusters in time $\Delta t_{1}$ and the old cluster population formed with the same star-formation time scale, $\delta t $, which depends on $\beta$ for consistency with the data in Figure (\ref{Fig6}). That is, here we assume $\Delta t_{2}=0$ and that the whole galaxy including GCs formed during $\Delta t_{1}$ (i.e $\Delta t_{1}=\Delta T$). From this assumption for each galaxy in our sample we calculate the maximum masses of the old cluster systems at a given SFR (equation \ref{eq14}). In this model the SFR is supposed to be constant over different time scales, i.e.,  for $\delta t$, $\Delta t_{1} $.

\begin{figure*}
    \centering
    \begin{subfigure}[]{0.1\textwidth}
        \centering
        \includegraphics[scale=0.65, angle=0]{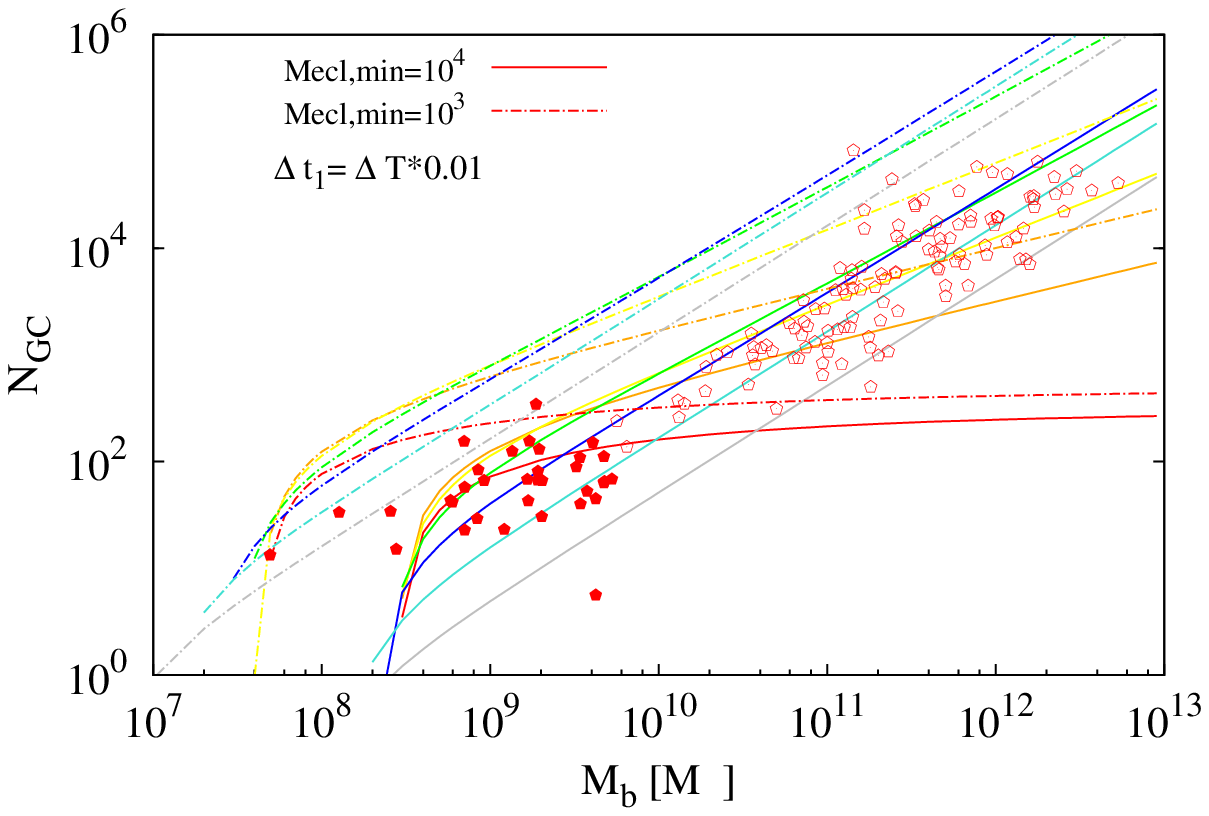}
        \caption*{}
        \label{Fig8a}
    \end{subfigure}
    \hfill
    \begin{subfigure}[]{0.5\textwidth}
        \centering
        \includegraphics[scale=0.65, angle=0]{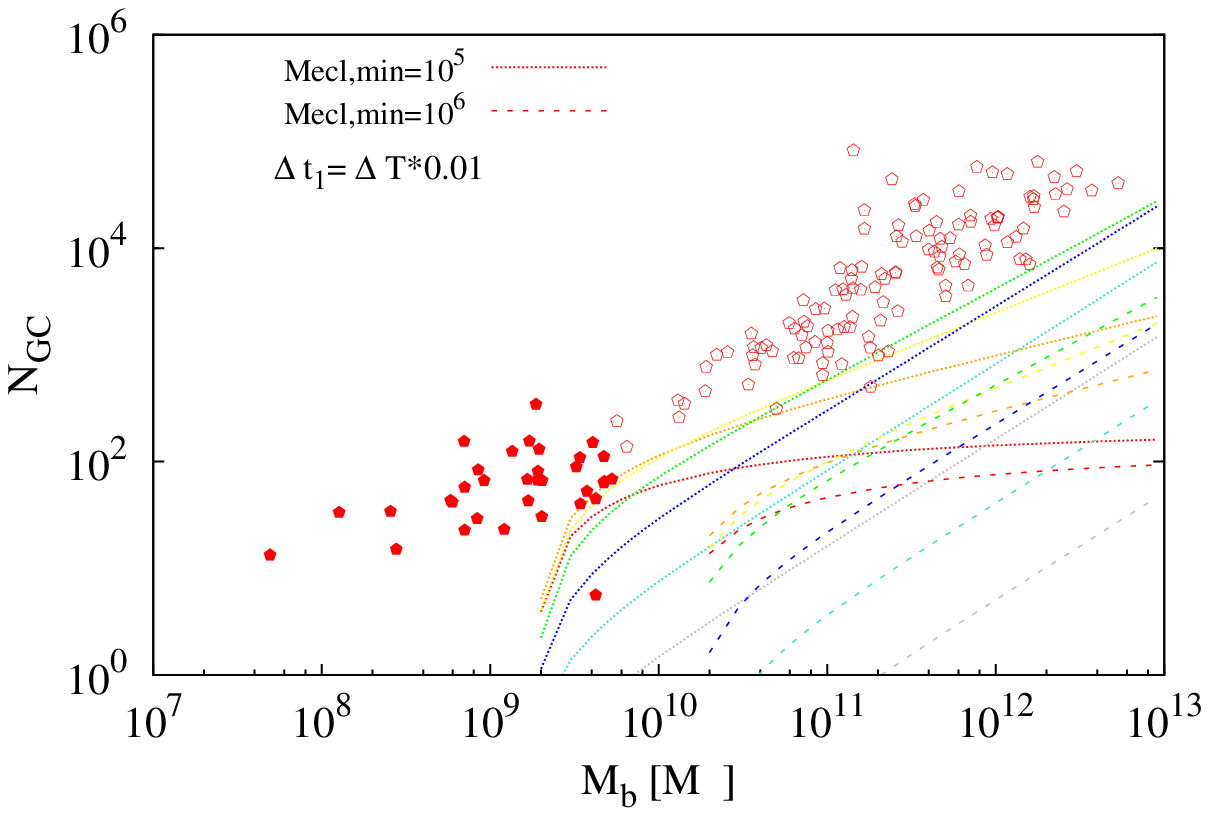}
        \caption*{}
        \label{Fig8b}
    \end{subfigure}
 \caption{Same as Figure (\ref{Fig9}) but for a model with $\Delta t_{1}= \Delta T \times 0.01$.}     
    \label{Fig10}
\end{figure*}

Having obtained $N_{GC,th}$ and  $\rm {M}_{b}$ and using $\psi$ in the $V$- band from equation (\ref{eq16}), we can compute $S_{N,th}$. By correcting the observed $N_{GC}$ (equation \ref{eq7*}) and $S_{N}$ (equation \ref{eq5})  for the erosion of GCs through tidal action or through dynamical friction, we obtain an estimate of the primordial values for each galaxy in our sample (as in Section 2).

\subsubsection{Comparison between the theoretical model and primordial value of {$N_{GC\MakeLowercase{i}}$} and  {$S_{N \MakeLowercase{i}}$}}

We investigate the influence of the two parameters $\beta$ and $ \rm {M}_{ecl,min}$ on  $N_{GCi}$ and $S_{Ni}$ for each galaxy. For this purpose, we calculate the model for seven values of $\beta$ (1.2, 1.5, 1.7, 1.9, 2.1, 2.3 and 2.5), and for four different values of $ \rm {M}_{ecl,min}$ ($10^{3}$,  $10^{4},  10^{5} $ and  $10^{6}\msun$). 
Since the overall distribution of $S_{Ni,iso}$ and $S_{Ni,aniso}$ is similar \citep{2014A&A...565L...6M}, we present the model only for the isotropic case. \newline
\begin{figure*}
\begin{center}
\includegraphics[width=150mm]{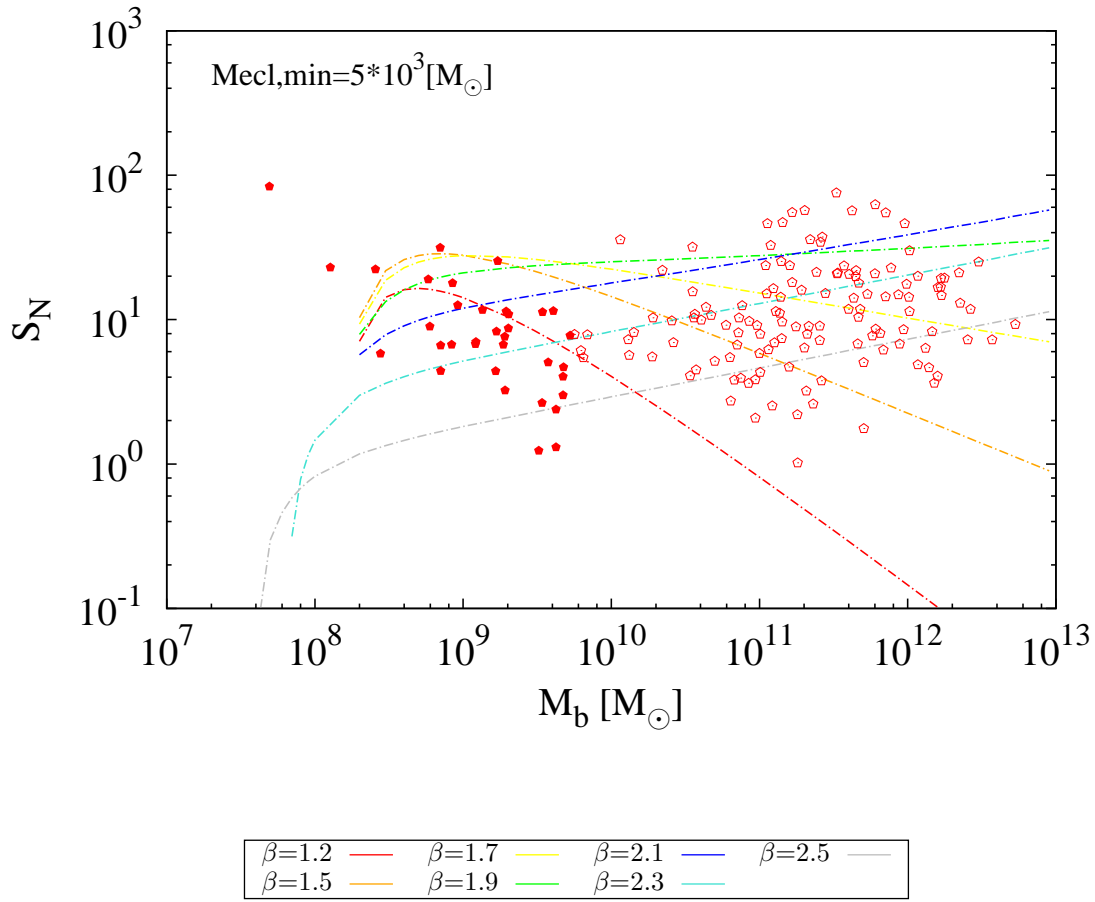}
\input{file-key.tex}
\caption{Comparison between the primordial value of the specific frequency of globular clusters and the theoretical specific frequency  for $\rm {M}_{ecl,min}=5 \times{10^{3}}\msun$  at $\Delta t_{1}= \Delta T \times {0.005}$ and $\Delta t_{2}=\Delta T- \Delta t_{1}$. The lines indicate our model for different $\beta$. The symbols are as in Figure (\ref{Fig9}).}
\label{Fig11}
\end{center}
\end{figure*}

The E and dE galaxies formed under different physical boundary conditions \citep{2000ApJ...543..149O,2013MNRAS.429.1858D}, which need different formation time-scales. We now also assume two values for ${\Delta t_{1}}$: firstly we assume ${\Delta t_{1}}$ to be equal to ${\Delta T}$ and secondly we assume it to be ${\Delta T} \times 10^{-2}$. This is to represent the formation epoch of the GC population which is likely to have been much shorter than the assembly time of the entire galaxy. 
Figure (\ref{Fig9}) shows the comparison between the primordial number of globular clusters ($ N_{GCi,iso}$) and the theoretical number of globular clusters ($N_{GC,th}$) for different $ \beta$ and $\rm {M}_{ecl,min}$ and for $\Delta t_{1} =\Delta T$, $\Delta t_{2}$ =0. The model does not represent  the observational data well for small $\rm {M}_{ecl,min}$, but agrees well with larger $\rm {M}_{ecl,min}$  $\geq 10^{5} \msun$ and only for galaxies in B\rm{II}. On the other hand, by using a smaller ${\Delta t_{1}}$, $\Delta t_{1}=\Delta T \times 0.01$, we match the observational data in B\rm{I} and B\rm{II} at lower $\rm {M}_{ecl,min}$ (Figure \ref{Fig10}). Thus, from Figures (\ref{Fig9}) and (\ref{Fig10}) we conclude that solutions are degenerate, the model does not need to be fine-tuned to account for the data.  

In Figure (\ref{Fig11}), we present a model for $S_{N,th}$ to match $S_{N,i}$ by setting $\rm {M}_{ecl,min}$ to be $5 \times 10^{3}\msun$  with $\Delta t_{1}=\Delta T \times 0.005$. It follows that dE galaxies are best represented by a model in which their GC population formed on a time scale $\Delta t_{1} \approx 0.005 \Delta T $ with $\rm {M}_{ecl,min} \approx 10^{3}\msun$ and $\beta\approx 1.2$. E galaxies require a  similar short time for the formation of their GC population but $\rm {M}_{ecl,min} \gtrsim  5 \times 10^{3}\msun$ and $\beta\approx 2.3$. Thus, the dE galaxies may have formed their GC population with a somewhat top-heavy ECMF, while massive star-bursting galaxies had an approximately Salpeter ECMF.  However, this conclusion is not unique, because solution to the dE galaxies with $\beta\approx 2.3$ are also possible by increasing the ratio  $\dfrac{\Delta t_{1}}{\Delta T}$.

\subsection{SFR not equal over $\Delta t_{1} $ and $\Delta t_{2} $}
\begin{figure}
\begin{center}
\includegraphics[scale=0.65, angle=0]{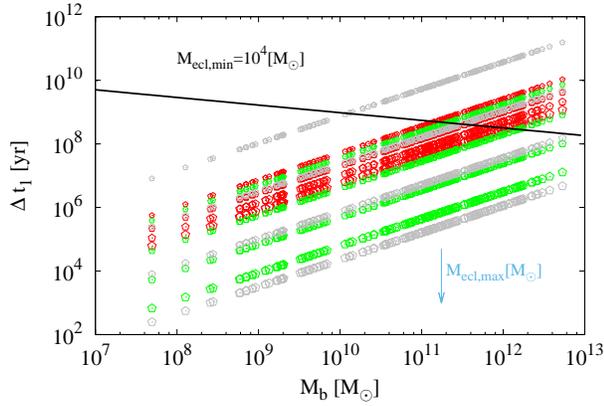}
\caption{The time of formation of the GC system, $\Delta t_{1}$, versus the total mass of a galaxy for $\rm {M}_{ecl,min}$=$10^{4}\msun$. The solid line is the time for forming the whole galaxy, $\Delta T$, from \cite{2009A&A...499..711R}. The colored points are for different $\beta$ of the ECMF $\beta$= 1.2, 1.9 and 2.5 (red, green and gray). Increasing symbol size indicates a higher maximum cluster mass.}
\label{Fig12}
\end{center}
\end{figure}

\begin{figure}
\begin{center}
\includegraphics[scale=0.65, angle=0]{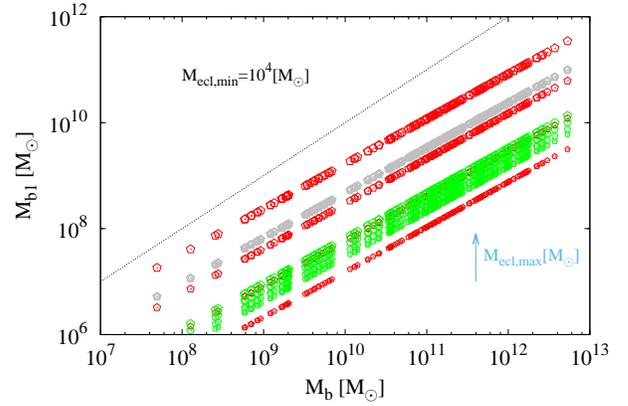}
\caption{The mass, $\rm {M}_{b1}$, of a galaxy formed during $\Delta t_{1}$ versus the total mass of a galaxy for $\rm {M}_{ecl,min}$=$10^{4}\msun$. The dotted line indicates  the 1:1 line. The color points and symbol sizes are the same as in Figure (\ref{Fig12}).}
\label{Fig13}
\end{center}
\end{figure}

In the following we compute the initial or primordial value of the number of GCs as a function of galaxy mass. As shown in  Figure (\ref{Fig5}), the SFRs need not be equal in the time  $\Delta t_{1}$ (SFR$_{1}$) and $\Delta t_{2}$ (SFR$_{2}$). That is we assume SFR$_{1}$ $\neq$ SFR$_{2}$, but the SFR to be constant within time span $\Delta t_{1}$ and $\Delta t_{2}$. By using equation (\ref{eq18}) and the observational data (Figure \ref{Fig3}) for $ N_{GCi,iso}$, we can estimate  $\Delta t_{1}$. We set the minimum star cluster mass equal to $10^{4}\msun$, as \cite{2003MNRAS.340..227B} suggested this as the minimum mass remaining bound as a cluster after 13 Gyr. We calculate the time scale $\Delta t_{1}$  for  $\rm {M}_{ecl,max}$ ranging between $10^{5}$ and $10^{8}\msun$ and for different $\beta$ for clarity, we display only $\beta$= 1.2, 1.9 and 2.5, see Figure (\ref{Fig12}). The solid black line indicates the star formation duration, $\Delta T$, as defined by equation (\ref{eq7}). Above this line, solutions become unphysical. $\rm {M}_{ecl,max}$ increases with decreasing $\Delta t_{1}$ and the difference of the models for different $\Delta t_{1}$ increases with $\beta$.
We calculate $\rm {M}_{b1}$ ($\rm {M}_{b1}= \rm {M}_{tot,\delta t}\times \frac{\Delta t_{1} }{\delta t}$) and compare these results to $\rm {M}_{b}$. In Figure (\ref{Fig13}) we directly compare the total mass which forms in  $\Delta t_{1}$ and the total mass of a galaxy. $\rm {M}_{b1}$ becomes unphysical above the dotted line, because larger than the total galaxy mass $\rm {M}_{b}$. As expected $\rm {M}_{b1}$ (total mass of stars formed during the GC formation epoch $\Delta t_{1}$)  is smaller than $\rm {M}_{b}$  (mass of galaxy).

\begin{figure}
\begin{center}
\includegraphics[scale=0.7, angle=0]{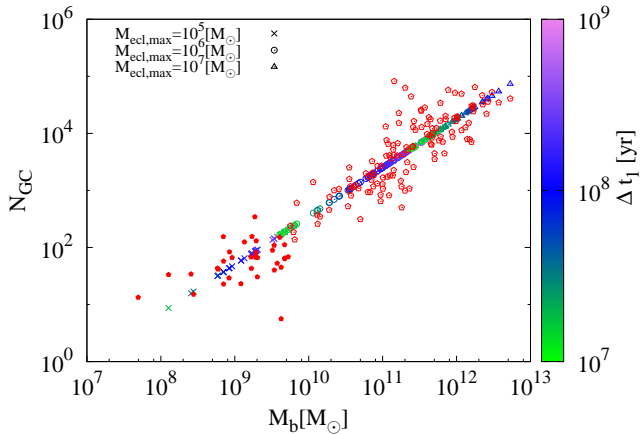}
\caption{Comparison between the observationally derived primordial value of the number of globular clusters ($N_{GC,iso}$; filled and open pentagons) and the theoretical number of globular clusters ($N_{GC,th}$) for  $\rm {M}_{ecl,min}$=$10^{4}\msun$ and $\beta$ = 2.3. Model symbols stand for different $\rm {M}_{ecl,max}$: crosses, $10^{5}\msun$; circles, $10^{6}\msun$; triangles, $10^{7}\msun$. The colour scale indicates different $\Delta t_{1}$.}
\label{Fig14}
\end{center}
\end{figure}

\begin{figure}
\begin{center}
\includegraphics[scale=0.7, angle=0]{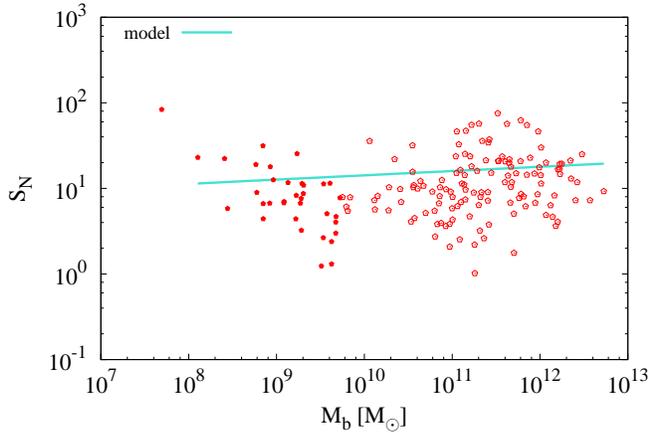}
\caption{The line is the theoretical specific frequency as a function of total galaxy mass ($\rm {M}_{b}$) for $\beta= 2.3$ (equation \ref{eq15}). In contrast with Figure (\ref{Fig11}), here $\rm {M}_{ecl,min}$=$10^{4}\msun$ and $\rm {M}_{ecl,max}$ varies between $10^{5}$ and $10^{7}\msun$ (see Section 4.2 for details). Symbols as in Figure (\ref{Fig9}).}
\label{Fig15}
\end{center}
\end{figure}

\cite{2004MNRAS.350.1503W} suggested a star-cluster mass function population
time scale ${\delta t}$ of about $10^{7}$ yr. Using this and the galaxy formation  time scale $\Delta T$  from downsizing \citep{2009A&A...499..711R}, we set $\Delta t_{1}$ between $10^{7}$ and $5\times 10^{8}$ yr. This is to obtain a physically realistic time scale for the formation of the GC system. By using equation (\ref{eq18}), we obtain  $N_{GC,th}$ depending on $\rm {M}_{ecl,max}$ and $\Delta t_{1}$. In Figure (\ref{Fig14}) we compare the primordial and theoretical values of $N_{GC,th}$ for $\rm {M}_{ecl,min} =10^{4} \msun$ and $\beta$ = 2.3 (since only $\beta$ equal to  2.3 can match $N_{GC,iso}$ for Branch \rm{I} and  \rm{II}). Figure (\ref{Fig14}) indicates that $\rm {M}_{ecl,max}$ increases with increasing $N_{GC,iso}$, and $\Delta t_{1} \approx 10^{8}$ yr represents most of the  $N_{GC,iso}$. The $S_{N,th}$ can be estimated using equation (\ref{eq15}) after obtaining $N_{GC,th}$ (Figure \ref{Fig15}). This model is thus able to account for the observed variation of $S_{N}$ with $\rm {M}_{b}$ for reasonable physical parameters.


\section{Summary and conclusions}
The specific frequency of GCs is a basic parameter to describe the GC system of a galaxy but remains poorly understood in the presently accepted galaxy formation framework \citep[see also][]{2015CaJPh..93..169K}. 
The overall trend indicates  high values of $S_{N}$ at opposite ends of the galaxy mass scale, while for a galaxy mass of around  $10^{10} \msun $, $S_{N}$ becomes close to one. The idea developed here extends the notion raised by \cite{2014A&A...565L...6M} to explain the `U' -shaped relation between  $S_{N}$ and $\rm {M}_{b}$ through tidal erosion. The most important driver of the erosion process is a tidal field. We support this idea  by showing the correlation between  $S_{N}$ and $\rho_{3D}$(Figure \ref{Fig2}). 
The GC survival fraction depends linearly on ${\rm log_{10}}$($\rho_{3D}$),  i.e.,  the observational data suggest that more GCs get destroyed in galaxies with a higher density which then have a smaller value of  $S_{N}$. It emerges that $S_{Ni}$ started approximately independently of  galaxy mass, $\rm {M}_{b}$, but later changed to a `U' -shape as a result of tidal erosion, which suggests that all early type galaxies had nearly the same efficiency to form young GCs. This in turn indicates that globular cluster formation is merely part of the universal star formation physics, which is evident in the solar neighborhood, and that it does not necessarily depend on the properties of the putative dark matter halos \citep[see also][]{2015CaJPh..93..169K}.

The primordial number of clusters for the $\textit{f}_{s,aniso}$ case is higher than that of $\textit{f}_{s,iso}$, because the erosion rate of GCs depends on the degree of the radial velocity anisotropy of the GC system \citep{2014MNRAS.441..150B}. \newline
We constructed a model to explain the initial specific frequency of GCs in galaxies at constant SFR during different time scales $\Delta t_{1}$ and $\Delta t_{2}$. 
A model is suggested according to which a population of young clusters is formed following a cluster mass function which depends on the SFR. The theoretical specific frequency of the GC model explains the primordial value of $S_{Ni}$, depending on the minimum star cluster mass and the slope of the cluster mass function. The models are  reasonably well fit to $S_{Ni}$ for $\rm {M}_{ecl,min}=5 \times 10^{3}\msun$ and $\Delta t_{1}=\Delta T \times 0.005$. According to the models, we can infer that for a low SFR (low galaxy mass) we need a lower minimum cluster mass and smaller $\Delta t_{1}$, while for a larger SFR (large galaxy masses) we need a higher minimum cluster mass. 

For the model with  $\Delta t_{1}$  shorter than $\Delta t_{2}$, we can match the primordial $S_{Ni}$ for $\Delta t_{1} \approx 10^{8}$ yr and  $\rm {M}_{ecl,min}$= $10^{4}\msun$. The best explanation for dE galaxies is the model with $\rm {M}_{ecl,max}$ = $10^{5} \msun$  and $\beta$ = 2.3. The best model for E galaxies is for $\rm {M}_{ecl,max}$ between  $10^{6}$ and $10^{7}$ $\msun$ and $\beta$ between 1.5 and 2.5. The existence of this difference may indicate a different formation mechanism for dE and E galaxies, respectively \citep{2000ApJ...543..149O,2013MNRAS.429.1858D}. We also see a possible hint that the embedded cluster mass function may become top-heavy (smaller $\beta$) in major galaxy-wide star burst, in support of the independent evidence found by \cite{2013MNRAS.436.3309W}
 
Thus, by considering that all stars form in correlated star formation events (i.e. embedded clusters) it is naturally possible to account for the observed dependency of $S_{N}$ on  galaxy mass $\rm {M}_{b}$. The large spread of $S_{N}$ values at a given $\rm {M}_{b}$ and the difference of $N_{GC}$ with $\rm {M}_{b}$ for dE and E galaxies (Branch \rm{I} and Branch \rm{II}, respectively) suggest that the detailed star-formation events varied between these systems. But the overall $S_{N}$ can be understood in terms of the above assumption, that is, in terms of universal purely baryonic processes playing the same role in all systems.

\section*{Acknowledgments}
We thank  M. Kruckow, M. Brockamp, A. H. W. K\"{u}pper, M. Marks and A. Dieball  for useful discussions and suggestions  and Z. Randriamanakoto and  C. Schulz  for providing data. We used the publicly available data from W. E. Harris (http://physwww.mcmaster.ca/$\sim$harris/Databases.html).
R.A.G.L. acknowledges the financial support of DGAPA, UNAM (PASPA program and project IN108518).

\bibliographystyle{apj}
\bibliography{paper}


\label{lastpage}

\end{document}